\documentclass[aps,prl,twocolumn,groupedaddress]{revtex4}

\usepackage{amssymb}
\usepackage{epsfig} 
\usepackage{amsmath} 
\usepackage{graphicx} 
\usepackage{color}

\newcommand{\eq}[1]{Eq.~(\ref{#1})}
\newcommand{\eqs}[2]{Eqs.~(\ref{#1}) and (\ref{#2})}

\newcommand{\GeV}{\mathinner{\mathrm{GeV}}}

\def\l{\left}
\def\r{\right}

\newcommand{\beq}{\begin{equation}}
\newcommand{\eeq}{\end{equation}}
\newcommand{\bea}{\begin{eqnarray}}
\newcommand{\eea}{\end{eqnarray}}

\begin{document}


\title{Invisible Higgs Decay Width vs. Dark Matter  Direct Detection  \\ Cross Section 
in Higgs Portal Dark Matter Models }



\author{Seungwon Baek}
\email[]{sbaek@kias.re.kr}
\affiliation{School of Physics, KIAS, Seoul 130-722, Korea}

\author{P. Ko}
\email[]{pko@kias.re.kr}
\affiliation{School of Physics, KIAS, Seoul 130-722, Korea}

\author{Wan-Il Park}
\email[]{wipark@kias.re.kr}
\affiliation{School of Physics, KIAS, Seoul 130-722, Korea}


\date{\today}

\begin{abstract}
The correlation between the invisible Higgs branching ratio ($B_h^{\rm inv} $) vs. dark matter 
(DM) direct detection  ($\sigma_p^{\rm SI}$)  in Higgs portal DM models is usually 
presented in the effective field theory (EFT) framework. 
This is fine for singlet scalar DM, but not in the singlet fermion DM (SFDM) or vector DM (VDM) models.  
In this paper, we derive the explicit expressions for this correlation within UV completions of 
SFDM and VDM models with Higgs portals, and discuss the limitation of the EFT approach. 
We show that there are at least two additional hidden parameter in $\sigma_p^{\rm SI}$ in the UV completions:  
the singlet-like scalar mass $m_2$ and its mixing angle $\alpha$ with the SM Higgs boson ($h$). 
In particular, if the singlet-like scalar is lighter than the SM Higgs boson ($m_2 < m_h \cos \alpha / \sqrt{1 + \cos^2 \alpha}$), the collider bound becomes weaker than the one based on EFT.
\end{abstract}


\maketitle


\section{Introduction}

As more data on the 126 GeV Higgs boson $H$ are accumulated at the LHC,  
its invisible Higgs branching fraction $B_h^{\rm inv}$ is getting bounded from above. 
This bound can give some useful constraint on the Higgs coupling to the DM particle 
in some concrete DM models.
In fact such attempts for Higgs portal DM  models were made recently by both ATLAS and CMS Collaborations~\cite{Aad:2014iia,Chatrchyan:2014tja}. Both Collaborations announced 
that their measurements of the upper bounds on the $B_h^{\rm inv} $ can be translated into 
the upper bounds on $\sigma_p$ (spin-independent cross section of DM particle on nucleon) 
in the Higgs portal DM models, which are much stronger than those obtained from DM direct 
detection experiments in the low DM mass region (i.e., $m_{\rm DM} \lesssim 10$ GeV).  
These analyses are based on the following model Lagrangians~\cite{Silveira:1985rk,Burgess:2000yq,Djouadi:2011aa,Djouadi:2012zc}:
\begin{widetext}
\begin{eqnarray}
{\cal L}_{\rm SSDM} &  = & \frac{1}{2} \partial_\mu S \partial^\mu S 
- \frac{1}{2} m_S^2 S^2 - \frac{\lambda_S}{4 !} S^4 
- \frac{\lambda_{HS}}{2} S^2 H^\dagger H
\\  \label{SFDM-EFT}
{\cal L}_{\rm SFDM} & = & \overline{\psi} ( i \partial - m_\psi ) \psi - 
\frac{\lambda_{\psi H}}{\Lambda} \overline{\psi} \psi H^\dagger H 
\\  \label{VDM-EFT}
{\cal L}_{\rm  VDM} & = &  - \frac{1}{4} V_{\mu\nu} V^{\mu\nu} + \frac{1}{2} m_V^2 
V_\mu V^\mu - \frac{\lambda_{VH} }{2} V_\mu V^\mu H^\dagger H 
- \frac{\lambda_V}{4} ( V_\mu V^\mu )^2
\end{eqnarray}
\end{widetext}
In all three cases, the DM phenomenology can be done with two parameters only, namely
the DM mass and the DM coupling to the Higgs field. The latter parameter is strongly
constrained by the upper bound on the invisible Higgs decay, and can be translated
into the upper bound on the spin-independent cross section of DM on nucleon.
This simple strategy has been adopted numerously.  

The SSDM Lagrangian (1) is renormalizable, and the results based on it would be reliable 
\footnote{There is an issue on the stability of DM using discrete $Z_2$ symmetry, which 
is beyond the scope this letter.}.
We refer to the comprehensive analyses on this model to the existing literature 
\cite{scalar_dm} without touching it in the following.
On the other hand, the other two cases for SFDM and VDM have to be considered
in better frameworks.  Since we don't know the new physics scales related with DM,  
we cannot know a priori how good the EFT approach would be.  Also the mass for the VDM 
is given by hand, so that both Lagrangians for SFDM and VDM are not renormalizable and 
violate unitarity at some scale.  In such cases, it is safer to consider simple UV completions of 
these two cases. 

In this letter, we point out that the claim by ATLAS and CMS based on the EFT 
is erroneous for SFDM and VDM cases,  by working in renormalizable and unitary Higgs portal 
DM models  proposed by the present authors~\cite{Baek:2011aa,Baek:2012se,Baek:2012uj}.  
In these two cases, there appears an additional SM singlet scalar, either from the renormalizable 
Yukawa couplings of the SFDM or from the remnant of dark Higgs mechanism for 
generating the VDM mass. In each case, we derive the expressions for the $B_h^{\rm inv}$ 
and $\sigma_p^{\rm SI}$,  and show that there are hidden variables in $\sigma_p^{\rm SI}$, 
namely the mass of the 
2nd scalar boson which is mostly singlet-like, and the mixing angle $\alpha$ between 
the SM Higgs and the singlet scalar boson. If kinematically allowed, the heavier scalar boson can decay into
a pair of lighter scalar bosons, so we have to consider the branching ratio for the 
nonstandard Higgs decays,  $B_h(m_h=125)^{\rm non SM}$ .  Then we use the LHC bounds on 
$B_h^{\rm inv}$ to derive the bounds on $\sigma_p^{\rm SI}$ as functions of 
$(m_2 , \alpha)$, and show when we recover the usual results presented by ATLAS and
CMS, and when we do not. This exercise will be not only physically important, but also
make good examples about the difference between the EFT and the full theory, and 
we would be able to understand clearly when the EFT can fail. 

In the following, we do not address thermal relic density of DM, since it is independent
of the issues raised and resolved in this paper.  It would be straightforward to include 
the discussions on thermal relic density, which would be presented elsewhere
~\cite{update}.


\section{Renormalizable  SFDM model}

The simplest renormalizable Lagrangian for the Higgs portal SFDM model is given by 
\cite{Baek:2011aa,Baek:2012uj} 
\footnote{One can consider another type of UV completion by introducing a new electroweak lepton doublet with the same quantum number as the SM Higgs doublet. 
The model is not a Higgs portal DM model, but is interesting in its own.  Detailed study of this model will be presented in a separate publication.}
\begin{widetext}
\begin{eqnarray} \label{sfdm}
{\cal L}_{\rm SFDM} & = & \overline{\psi} \left( i \partial - m_\psi - \lambda_\psi S \right)
- \mu_{HS} S H^\dagger H - \frac{\lambda_{HS}}{2} S^2 H^\dagger H
\nonumber \\
& + & \frac{1}{2} \partial_\mu S \partial^\mu S - \frac{1}{2} m_S^2 S^2 - \mu_S^3 S 
- \frac{\mu_S^{'}}{3} S^3 - \frac{\lambda_S}{4} S^4 .
\end{eqnarray}
\end{widetext}
We consider Dirac fermion DM in this paper. 
For the Majorana fermion DM case, we have to multiply a factor $1/2$ to the \textit{invisible} decay rate of Higgses, and it results in a factor $2$ larger $\sigma_p^{\rm SI}$ relative to the case of Dirac fermion DM.
In general, the singlet scalar $S$ can develop a nonzero VEV, and we have to shift 
the field as $S (x) \rightarrow \langle S \rangle + s(x)$. 
Also the SM Higgs will break the EWSB spontanesouly.  The detailed expressions for the relations among 
various parameters can be found in Ref.~\cite{Baek:2011aa}, to which we refer the 
details. 

After all, there are two scalar bosons, a mixture of the SM Higgs boson $h$ and the singlet 
scalar $s$.  The physical states are defined after the $SO(2)$ rotation:
\begin{eqnarray*}
H_1 & = & h \cos\alpha - s \sin \alpha , \\ 
H_2 & = & h \sin \alpha + s \cos \alpha .
\end{eqnarray*}
Note that there is a minus sign in one term which orginates from $SO(2)$ nature of the rotation 
matrix in the scalar sector.  This minus sign plays an important role
in the direct detection cross section of the DM scattering on nucleon, since the contributions of 
$H_1$ and $H_2$ to $\sigma_p$ interferes destructively~\cite{Baek:2011aa}. 
This  is a very generic phenomenon in both SFDM and VDM cases~\cite{Baek:2011aa,Baek:2012se}
\footnote{This phenomenon is similar to the GIM mechanism in the quark flavor physics.}.

The invisible and the non-SM branching fractions of Higgs decay  and the DM-proton scattering cross section 
within the renormalizable SFDM model are given as follows: 
\begin{eqnarray}
\label{Bi-inv}
B_i^{\rm inv} & = & \frac{\l( 1 - \kappa_i(\alpha) \r) \Gamma_i^{\rm inv}}{\kappa_i(\alpha) 
\Gamma_i^{\rm SM} + \l( 1 - \kappa_i(\alpha) \r) \Gamma_i^{\rm inv} + \Gamma_i^{jj}}
\\
\label{Bi-non-sm}
B_i^{\rm non SM} & = & \frac{\Gamma_i^{jj}}{\kappa_i(\alpha) 
\Gamma_i^{\rm SM} + \l( 1 - \kappa_i(\alpha) \r) \Gamma_i^{\rm inv} + \Gamma_i^{jj}}
\\
\label{sp-sfdm}
\sigma_p^{\rm SI} & = & \frac{m_r^2}{\pi} \l( \frac{\lambda_\psi s_\alpha c_\alpha \ m_p}{v_H} \r)^2 \mathcal{F}(m_\psi, \{m_i\},v) f_p^2
\end{eqnarray}
where $\kappa_i(\alpha) = c_\alpha^2, s_\alpha^2$ for $i=1,2$, the decay rates of Higgs particles are given by 
\bea
\Gamma_i^{\rm SM} &=& \Gamma_h (m_i)
\\
\Gamma_i^{\rm inv} &=& \frac{\lambda_\psi^2}{8 \pi} m_i \l( 1 - \frac{4 m_\psi^2}{m_i^2} \r)^{3/2}
\\ \label{Gammaijj}
\Gamma_i^{jj} &=& \frac{1}{32 \pi m_i} \lambda_{ijj}^2 \l( 1 - \frac{4 m_j^2}{m_i^2} \r)^{1/2}
\eea
with $\lambda_{ijj}$ which is given by
\bea
\lambda_{122} 
&=& \lambda_{HS} v_H c_\alpha^3 + 2 \l( 3 \lambda_H - \lambda_{HS} \r) v_H c_\alpha s_\alpha^2 
 \\
&& - 2 \l[ \mu_S' +3 \l( \lambda_S - \lambda_{HS} \r) v_S \r] c_\alpha^2 s_\alpha - \lambda_{HS} v_S s_\alpha^3
\nonumber
\\
\lambda_{211} 
&=& \lambda_{HS} v_S c_\alpha^3 + 2 \l( 3 \lambda_H - \lambda_{HS} \r) v_H c_\alpha^2 s_\alpha
 \\
&& 2 \l[ \mu_S' +3 \l( \lambda_S - \lambda_{HS} \r) v_S \r] c_\alpha s_\alpha^2 + \lambda_{HS} v_H s_\alpha^3
\nonumber
\eea
and 
\bea
\mathcal{F}
&=& \frac{1}{4 m_\psi^2 v^2} \l[ \sum_i \l( \frac{1}{m_i^2} - \frac{1}{4 m_\psi^2 v^2 + m_i^2} \r) \r.
\\
&& \l. - \frac{2}{\l( m_2^2 - m_1^2 \r)} \sum_i \l(-1\r)^{i-1} \ln \l( 1 + \frac{4 m_\psi^2 v^2}{m_i^2} \r) \r]
\nonumber 
\eea 
with $v$ being the lab velocity of DM, and $m_r \equiv m_\psi m_p / \l( m_\psi + m_p \r)$ and $f_p = \sum_{q=u,d,s} f_q + \frac{2}{9} f_Q$ with $f_q$ being the hadronic matrix element and 
$f_Q = 1-\sum_{q=u,d,s} f_q$.  We take the $f_p = 0.326$ from a lattice calculation \cite{Young:2009zb}. 
Note that the channel,``$h \to \phi \phi^* \to \phi b \bar{b}$'' is also possible, and the associated decay rate is
\beq
\Gamma_{h \to \phi b \bar{b}} \sim \frac{\l( \lambda_{122} s_\alpha \r)^2 }{3 \l( 2 \pi \r)^5} \l( \frac{m_b}{m_h} \r)^2 \frac{\l( m_h - m_\phi \r)^5}{m_h m_\phi^5}
\eeq
This is smaller than $\Gamma_h^{\rm SM}$ by many orders of magnitude, and can be ignored safely.
\begin{figure}[h]
\centering
\includegraphics[width=0.45\textwidth]{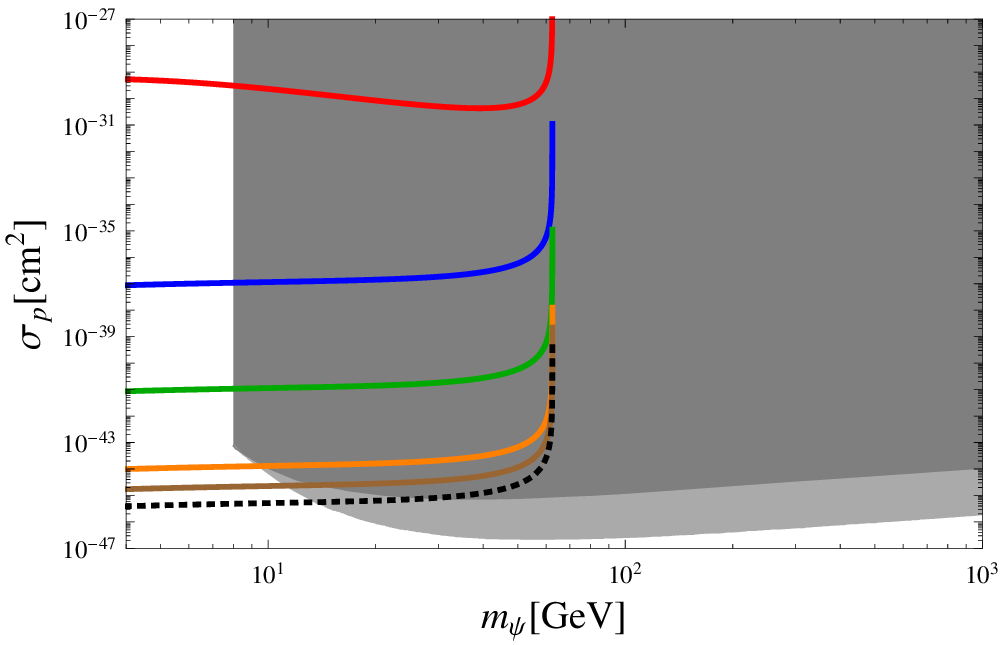}
\includegraphics[width=0.45\textwidth]{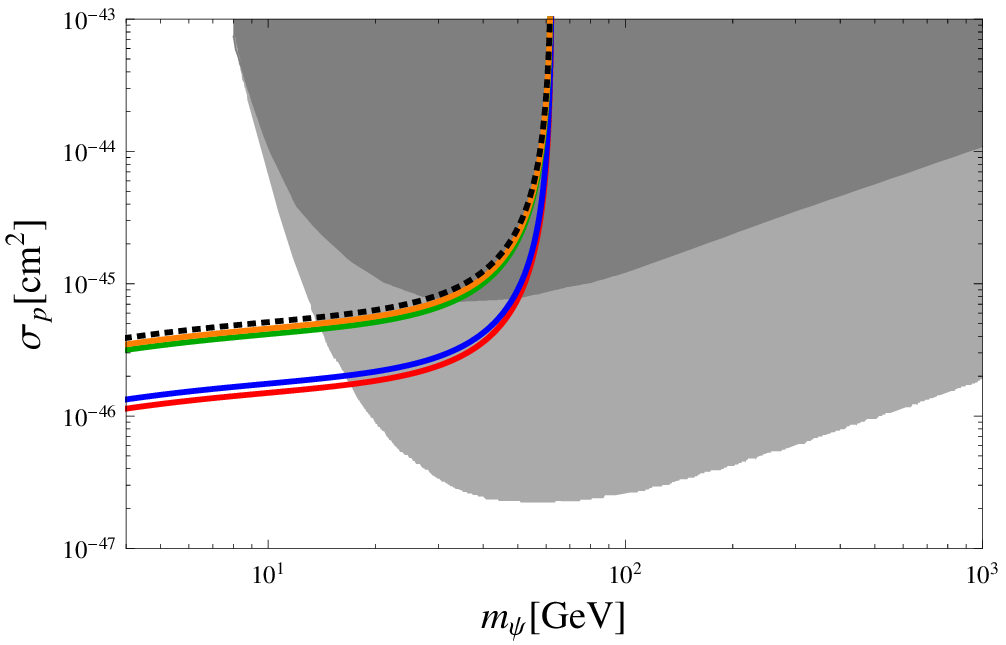}
\caption{\label{fig:sp-SFDM} $\sigma_p^{\rm SI}$ as a function of the mass of dark matter for SFDM for a mixing angle $\alpha=0.2$.
Upper panel: $m_2 = 10^{-2}, 1, 10, 50, 70 \GeV$ for solid lines from top to bottom.
Lower panel: $m_2 = 100, 200, 500, 1000 \GeV$ for dashed lines from bottom to top.
The balck dotted line is EFT prediction. 
Dark-gray and gray region are the exclusion regions of LUX \cite{Akerib:2013tjd} and projected XENON1T (gray) \cite{Aprile:2012zx}.
\label{fig:sigmap}}
\end{figure}

Let us compare these results with those obtained in the EFT:
\begin{eqnarray}
\label{B1-eft}
( B_h^{\rm inv} )_{\rm EFT} & = & \frac{\l( \Gamma_h^{\rm inv} \r)_{\rm EFT}}{\Gamma_h^{\rm SM} + \l( \Gamma_h^{\rm inv} \r)_{\rm EFT}}
\\
\label{sp-sfdm-eft}
( \sigma_p^{\rm SI} )_{\rm EFT} & = & \frac{m_r^2}{\pi} \l[ \frac{\lambda_{\psi H}  \ m_p}{\Lambda m_h^2} \r]^2 f_p^2 
\end{eqnarray}
where
\beq
( \Gamma_h^{\rm inv} )_{\rm EFT} = \frac{1}{8 \pi} \l( \frac{\lambda_{\psi H} v_H}{\Lambda} \r)^2 m_h \l( 1 - \frac{4 m_\psi^2}{m_h^2} \r)^{3/2} .
\eeq
Recent analysises of LHC experiments impose a bound \cite{Aad:2014iia,Chatrchyan:2014tja} 
on the branching fraction of SM-like Higgs decay to invisible particles as \cite{Chatrchyan:2014tja}
\beq
B_h^{\rm inv} < 0.51 \ {\rm at} \ 95\%{\rm CL} 
\eeq
(see also Ref.~\cite{Chpoi:2013wga} for more involved analysis in the presence of extra singlet-like 
scalar boson that mixes with the SM Higgs boson).  
In the renormalizable model described by \eq{sfdm}, the LHC bound on $B_h^{\rm inv}$ can be translated 
directly to a constraint on $\sigma_p^{\rm SI}$ by the relation, 
\bea
\sigma_p^{\rm SI} \label{sp-full}
&=& c_\alpha^4 m_h^4 \mathcal{F}(m_\psi, \{m_i\},v)
\nonumber \\
&\times&
\frac{B_h^{\rm inv} \Gamma_h^{\rm SM}}{\l( 1-B_h^{\rm inv} \r)} \frac{8 m_r^2}{m_h^5 \beta_\psi^3} \l( \frac{m_p}{v_H} \r)^2 f_p^2
\eea
where $\beta_\psi = \sqrt{1-4m_\psi^2/m_h^2}$.  
Here  we set $B_1^{\rm nonSM}=0$ for simplicity,  and denoted $B_1^{\rm inv}$ as $B_h^{\rm inv}$. 
On the other hand, in the EFT described by \eq{SFDM-EFT} with 
$\l( B_h^{\rm inv} \r)_{\rm EFT} \to {\rm B}_h^{\rm inv}$,  one finds
\beq \label{sp-eft}
\l( \sigma_p{\rm SI} \r)_{\rm EFT} = \frac{B_h^{\rm inv} \Gamma_h^{\rm SM}}{1- B_h^{\rm inv}} \frac{8 m_r^2}{m_h^5 \beta_\psi^3} \l( \frac{m_p}{v_H} \r)^2 f_p^2 
\eeq
which was used in the analysis's of ATLAS \cite{Aad:2014iia} and CMS \cite{Chatrchyan:2014tja}.
Now it is clear from \eqs{sp-full}{sp-eft} that, contrary to $\l( \sigma_p^{\rm SI} \r)_{\rm EFT}$ of EFT, 
$\sigma_p^{\rm SI}$ of a full theory of \eq{sfdm} has additional factors, $c_\alpha^4 m_h^4 {\cal F}$,  
which involves two extra parameters, ($\alpha$, $m_2$). 
Note that, in the limit $\alpha$ is very small so that we can make $\cos \alpha \simeq 1$, and $m_2 \gg m_1$ so that we can drop $1/m_2^2$ term in the $\sigma_p^{\rm SI}$,  \eq{sp-full} for $\sigma_p^{\rm SI}$ 
approaches to \eq{sp-eft} for $\l(\sigma_p^{\rm SI}\r)_{\rm EFT}$.  
However, if one of these two assumptions is not valid, one cannot make
a definitive prediction for the $\sigma_p^{\rm SI}$.  Therefore the bounds on the 
$\sigma_p^{\rm SI}$ derived by the ATLAS and the CMS Collaborations should be taken
with caution.  Basically one cannot make model-independent connections between
$B_h^{\rm inv} (=B_1^{\rm inv})$ and $\sigma_p^{\rm SI}$ in the Higgs portal SFDM model.  
This is clearly shown in Fig.~\ref{fig:sigmap} where colored solid lines represent the LHC bound on 
$\sigma_p^{\rm SI}$ of \eq{sp-sfdm} for various values for $m_2$.
The bound on $( \sigma_p^{\rm SI} )_{\rm EFT}$ of \eq{sp-sfdm-eft} was also depicted for comparison. 
Note that, for low $m_\psi$ if $m_2 < m_h c_\alpha / \sqrt{1 + c_\alpha^2}$, the LHC bound becomes weaker than the claims made in \cite{Aad:2014iia,Chatrchyan:2014tja}.
Especially, for $m_2 \lesssim m_h c_\alpha/ \sqrt{12.3 + c_\alpha^2}$, it can not win the direct detection bound for $m_\psi \gtrsim 8 \GeV$.

\section{Renormalizable VDM model}

The simplest renormalizable Lagrangian for the Higgs portal VDM model is given by 
\cite{Baek:2012se,Farzan:2012hh} 
\begin{widetext}
\begin{eqnarray} \label{vdm}
{\cal L}_{\rm VDM} & = & - \frac{1}{4} V_{\mu\nu} V^{\mu\nu} + D_\mu \Phi^\dagger
D^\mu \Phi - \lambda_{\Phi} \left( \Phi^\dagger \Phi - \frac{v_\Phi^2}{2} \right)^2 
- \lambda_{\Phi H} \left( \Phi^\dagger \Phi - \frac{v_\Phi^2}{2} \right) 
\left( H^\dagger H - \frac{v_H^2}{2} \right)
\end{eqnarray}
\end{widetext}
where $\Phi$ is the dark Higgs field which generates nonzero mass for the VDM
through spontaneous $U(1)_X$ breaking, and 
\[
D_\mu \Phi \equiv \left( \partial_\mu + i g_X Q_{\Phi} V_\mu \right) \Phi
\]
After $U(1)_X$ breaking, we shift the field $\Phi_X$ as follows:
\[
\Phi \rightarrow \frac{1}{\sqrt{2}} ( v_\Phi + \phi(x) )
\]
where the field $\phi(x)$ is a SM singlet scalar similarly to the singlet scalar in the 
SFDM case.  Again there are two scalar bosons which are mixtures of $h$ and $\phi$.

The invisible and non-SM branching fractions of the Higgs decay are of the same forms as \eqs{Bi-inv}{Bi-non-sm}, but with
\beq
\Gamma_i^{\rm inv} = \frac{g_X^2}{32 \pi} \frac{m_i^3}{m_V^2} \l( 1 - \frac{4 m_V^2}{m_i^2} + 12 \frac{m_V^4}{m_i^4} \r) \l( 1 - \frac{4 m_V^2}{m_i^2} \r)^{1/2}
\eeq
where $m_V$ is the mass of VDM, and $\Gamma_i^{jj}$ with $\mu_P'=0$.
The spin-indenpendent cross section of VDM to proton is also same as the one of \eq{sp-sfdm} with $\lambda_\psi$ and $m_\psi$ replaced to $g_X$ and $m_V$, respectively.

Again, let us compare these results with those in the EFT:
$\l({\rm B}_h^{\rm inv}\r)_{\rm EFT}$ is of the same form as \eq{B1-eft} with
\bea
( \Gamma_h^{\rm inv} )_{\rm EFT} 
&=& \frac{\lambda_{VH}^2}{128 \pi} \frac{v_H^2 m_h^3}{m_V^4} \times
\nonumber \\
&& \l( 1 - \frac{4 m_V^2}{m_h^2} + 12 \frac{m_V^4}{m_h^4} \r) \l( 1 - \frac{4 m_V^2}{m_h^2} \r)^{1/2} 
\eea
and the VDM-nucleon scattering cross section is 
\beq \label{sp-vdm-eft-limit}
( \sigma_p^{\rm SI} )_{\rm EFT} = \frac{m_r^2}{\pi} \l[ \frac{\lambda_{V H}  \ m_p}{2 m_V m_h^2} \r]^2 f_p^2  
\eeq
\begin{figure}[h]
\centering
\includegraphics[width=0.45\textwidth]{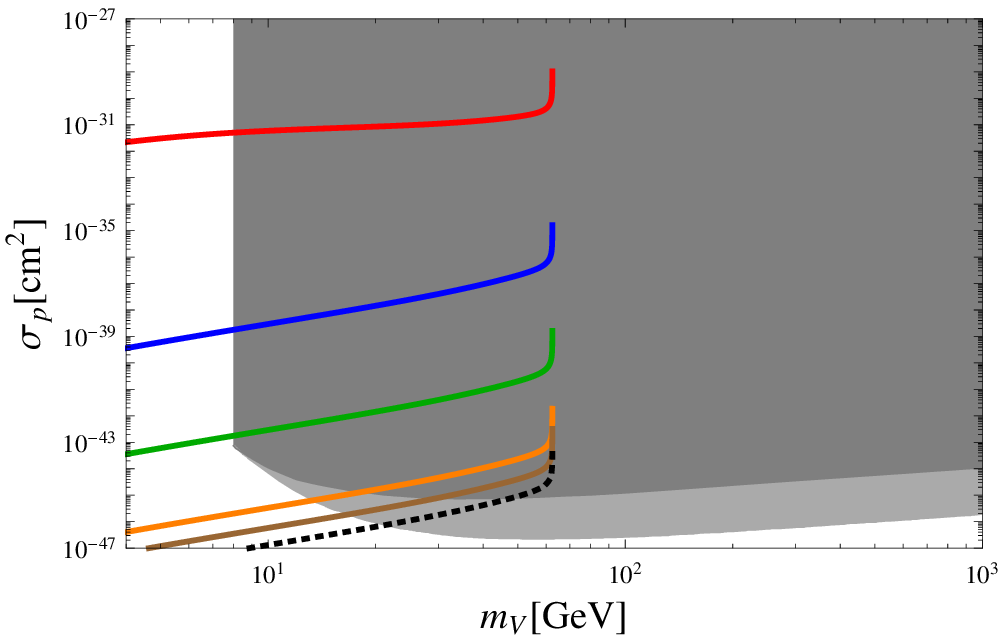}
\includegraphics[width=0.45\textwidth]{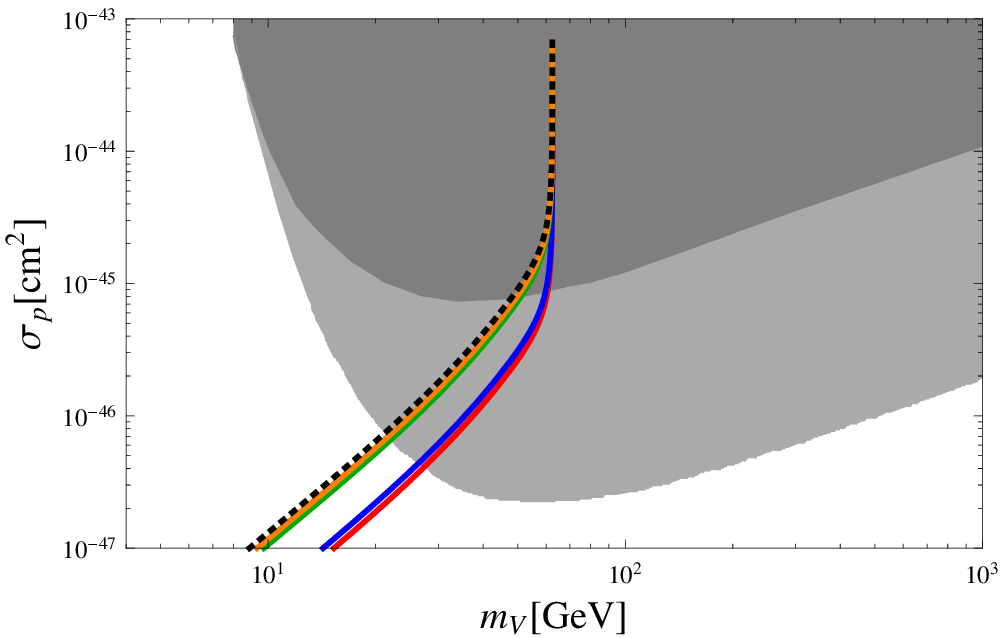}
\caption{\label{fig:sp-SVDM} $\sigma_p^{\rm SI}$ as a function of the mass of dark matter for SVDM for a mixing angle $\alpha=0.2$.
Same color and line scheme as Fig.~\ref{fig:sp-SFDM}.
}
\end{figure}
In the renormalizable model of \eq{vdm}, the LHC bound on $B_h^{\rm inv}$ can be translated directly to a constraint on $\sigma_p^{\rm SI}$ by the relation, 
\bea
\sigma_p^{\rm SI} \label{sp-vdm-full}
&=& c_\alpha^4 m_h^4 \mathcal{F}(m_V, \{m_i\},v)
\nonumber \\
&\times&
\frac{B_h^{\rm inv} \Gamma_h^{\rm SM}}{\l( 1-B_h^{\rm inv} \r)} \frac{32 m_r^2 m_V^2 \l( m_p /v_H \r)^2 f_p^2}{m_h^7 \beta_V \l( 1 - \frac{4 m_V^2}{m_h^2} + 12 \frac{m_V^4}{m_h^4} \r)}
\eea
where $\beta_V = \sqrt{1-4m_V^2/m_h^2}$.
On the other hand, in the EFT of \eq{VDM-EFT} one finds
\beq \label{sp-vdm-eft}
\l( \sigma_p^{\rm SI} \r)_{\rm EFT} = \frac{B_h^{\rm inv} \Gamma_h^{\rm SM}}{1- B_h^{\rm inv}} \frac{32 m_r^2 m_V^2 \l( m_p /v_H \r)^2 f_p^2}{m_h^7 \beta_V \l( 1 - \frac{4 m_V^2}{m_h^2} + 12 \frac{m_V^4}{m_h^4} \r)} 
\eeq
used in the analysis's of ATLAS \cite{Aad:2014iia} and CMS \cite{Chatrchyan:2014tja}.
Note again that $\sigma_p^{\rm SI}$ of \eq{sp-vdm-full} has additional factors involving ($\alpha$, $m_2$), compared to $\l(\sigma_p^{\rm SI} \r)_{\rm EFT}$ of \eq{sp-vdm-eft}.
Therefore, similarly to the case of SFDM, one cannot make model-independent connections between $B_h^{\rm inv}$ and $\sigma_p^{\rm SI}$ in the Higgs portal VDM model.  
Fig.~\ref{fig:sp-SVDM}, where $\sigma_p^{\rm SI}$ of \eq{sp-vdm-full} and $( \sigma_p^{\rm SI} )_{\rm EFT}$ of \eq{sp-vdm-eft} in VDM scenario are depicted for comparison, shows clearly this discrepancy caused by the different dependence on $\alpha$ and $m_2$.

\section{Implications for DM search and collider experiments}

From our arguments based on the renormalizable and unitary model Lagrangians,
it is clear that one has to seek for the singlet-like second scalar boson $H_2$.
It could be either lighter or heavier than the observed Higgs boson. Since the observed 
125 GeV Higgs boson has a signal strength $\sim 1$, the other ons has the signal strength 
$\lesssim 0.1$.  Therefore it would require dedicated searches for this singlet-like scalar boson 
at the LHC.
In fact this second scalar boson is almost ubiquitous in hidden sector DM models, where DM
is stabilized or long-lived due to dark gauge symmetries~\cite{hur_ko,Baek:2012se,Baek:2013qwa,Baek:2013dwa,Baek:2014goa,Ko:2014nha,Ko:2014bka}. 
In case this second scalar is light, it could solve some puzzles in the CDM paradigm, 
such as core cusp problem,  missing satellite problem or too-big-to-fail problem~
\cite{Ko:2014nha,Ko:2014bka}.  
And it can help the Higgs inflation work ~\cite{Ko:2014eia}  in light of the recent BICEP2 
results with large tensor-to-scalar ratio $r = 0.2^{+0.07}_{-0.05}$.
Therefore it would be very important to search for the singlet-like second scalar boson
at the LHC and elsewhere, in order to test the idea of dark gauge symmetry stabilizing
the DM of the universe. 
Since the ILC can probe $\alpha$ down to a few $\times 10^{-3}$ only, there would be 
an ample room for the 2nd scalar remaining undiscovered at colliders unfortunately. 
It would be a tough question how to probe the region below $\alpha \lesssim 10^{-3}$ 
in the future terrestrial experiments ( for example, see \cite{Buckley:2014ika} for a recent study). 

The second point is that there is no unique correlation between the LHC data 
on the Higgs invisible branching ratio and the spin-independent cross section of
Higgs portal DM on nucleon. One can not say that the former gives stronger bound
for low DM mass region compared with the latter, which is very clear from the plots we have shown.
Therefore it is important for the direct detection experiments to improve the upper 
bound on $\sigma_{\rm SI}$ for low $m_{\rm DM}$, regardless of collider bounds.
Collider bounds can never replace the DM direct search bounds in a model independent
way, unlike many such claims.

\section{Conclusion}

In this letter, we have demonstrated that the effective theory approach 
in dark matter physics could lead to erroneous or misleading results.  
For the Higgs portal SFDM and VDM, there are at least two more important parameters, 
the mass $m_2$ of the 2nd scalar which is mostly a SM singlet, and the mixing angle 
$\alpha$ between the SM Higgs boson and the 2nd scalar boson:
\begin{eqnarray}
\sigma_p^{\rm SI} & = & ( \sigma_p^{\rm SI} )_{\rm EFT}~c_\alpha^4  m_h^4 {\cal F} ( m_{\rm DM}, 
\{ m_i \} , v)
\\
& \simeq & ( \sigma_p^{\rm SI} )_{\rm EFT}~c_\alpha^4 \left(  1 - \frac{m_h^2}{m_2^2} \right)^2
\end{eqnarray}
where the function  ${\cal F}$ is defined in Eq.~(13) and $m_1 = m_h = 125$ GeV.  
The second equation is  obtained when the momentum of DM is negligible relative to both masses of Higgses.
The usual EFT approach applies only for the case $m_2 = m_h c_\alpha/\sqrt{1+c_\alpha^2}$ or $m_2 \rightarrow \infty$
with $\alpha \to 0$.  For the finite $m_2$, there is a generic cancellation 
between $H_1$ and $H_2$ contribution due to the orthogonal nature of the rotation 
matrix from interaction to mass eigenstates of two scalar bosons.
The resulting bound on $\sigma_{\rm SI}$ becomes even stronger if $m_2 > m_1 = 125$ GeV.
On the other hand, for a light 2nd Higgs ($m_2 < m_h c_\alpha / \sqrt{1 + c_\alpha^2}$), the LHC bound derived from the invisible  Higgs decay width is weaker than the claims made in both ATLAS and CMS collaborations.
Especially, for $m_2 \lesssim m_h c_\alpha/ \sqrt{12.3 + c_\alpha^2}$, it can not compete with the DM direct search bounds from XENON100, CDMS  and LUX, which is the main conclusion of this paper.
Both LHC search for the singlet-like 2nd scalar boson and the DM direct search 
experiments are important to be continued, and will be complementary with each other.

\begin{acknowledgments}
We thank Suyong Choi, Teruki Kamon, Sungwon Lee and Un-Ki Yang for useful 
discussions on the subject presented in this work. 
This work was supported in part by Basic Science Research Program through the
National Research Foundation of Korea (NRF) funded by NRF Research Grant 
2012R1A2A1A01006053 (SB,PK,WIP), and by SRC program of NRF funded by 
MEST (20120001176) through Korea Neutrino Research Center at Seoul National
University (PK).  
\end{acknowledgments}


\end{document}